\pdfoutput=1
\documentclass[onecolumn,showpacs,amsmath,amssymb,12pt]{revtex4}
\usepackage{graphicx}
\usepackage{dcolumn}
\usepackage{bm}
\begin{document}
\newcommand{\hs}{\hspace*{0.5cm}}
\newcommand{\vs}{\vspace*{0.5cm}}
\newcommand{\be}{\begin{equation}}
\newcommand{\ee}{\end{equation}}
\newcommand{\bea}{\begin{eqnarray}}
\newcommand{\eea}{\end{eqnarray}}
\newcommand{\ben}{\begin{enumerate}}
\newcommand{\een}{\end{enumerate}}
\newcommand{\bde}{\begin{widetext}}
\newcommand{\ede}{\end{widetext}}
\newcommand{\nn}{\nonumber}
\newcommand{\crn}{\nonumber \\}
\newcommand{\Tr}{\mathrm{Tr}}
\newcommand{\non}{\nonumber}
\newcommand{\noi}{\noindent}
\newcommand{\al}{\alpha}
\newcommand{\la}{\lambda}
\newcommand{\bet}{\beta}
\newcommand{\ga}{\gamma}
\newcommand{\va}{\varphi}
\newcommand{\om}{\omega}
\newcommand{\pa}{\partial}
\newcommand{\+}{\dagger}
\newcommand{\fr}{\frac}
\newcommand{\bc}{\begin{center}}
\newcommand{\ec}{\end{center}}
\newcommand{\Ga}{\Gamma}
\newcommand{\de}{\delta}
\newcommand{\De}{\Delta}
\newcommand{\ep}{\epsilon}
\newcommand{\varep}{\varepsilon}
\newcommand{\ka}{\kappa}
\newcommand{\La}{\Lambda}
\newcommand{\si}{\sigma}
\newcommand{\Si}{\Sigma}
\newcommand{\ta}{\tau}
\newcommand{\up}{\upsilon}
\newcommand{\Up}{\Upsilon}
\newcommand{\ze}{\zeta}
\newcommand{\ps}{\psi}
\newcommand{\Ps}{\Psi}
\newcommand{\ph}{\phi}
\newcommand{\vph}{\varphi}
\newcommand{\Ph}{\Phi}
\newcommand{\Om}{\Omega}
\newcommand{\Huong}[1]{{\color{blue}#1}}
\newcommand{\Duy}[1]{{\color{green}#1}}

\title{Investigation of the Higgs boson anomalous FCNC interactions in the simple 3-3-1 model }
\author{D. T. Huong$^{a}$}
\email{dthuong@iop.vast.ac.vn}
\author{N. T. Duy $^{a,b}$}

\affiliation{
	$^a$ Institute of Physics, VAST, 10 Dao Tan, Ba Dinh, Hanoi, Vietnam\\
	$^b$ Graduate University of Science and Technology,
	Vietnam Academy of Science and Technology,
	18 Hoang Quoc Viet, Cau Giay, Hanoi, Vietnam}
\email{ntdem@iop.vast.ac.vn}
\date{\today}

\begin{abstract}
We study phenomenological constraints on the simple 3-3-1 model with flavor-violating Yukawa couplings. Both Higgs triplets couple to leptons and quarks, which generates flavor-violating signals in both lepton and quark sectors. We have shown that this model allows for a large Higgs lepton flavor-violating rate decay $h \rightarrow \mu \tau$ and may reach perfect agreements with other experimental constraints such as $\tau \rightarrow \mu \gamma$ and $(g-2)_\mu$. The contributions of flavor-changing neutral current couplings, Higgs--quark--quark couplings, mixing to the mesons are investigated. Br$(h \rightarrow q q^\prime )$ can be enhanced acknowledging the measurements of meson mixing. The branching ratio for $t \rightarrow q h$  can reach up to $10^{-3}$, but it could be as low as $10^{-8}$. 

\end{abstract}


\maketitle

\section{\label{intro}Introduction}

The discovery of the Higgs boson in July 2012 at the Large Hadron Collider (LHC) \cite{atlascms} has opened up a new area of the direct search for physics beyond the Standard Model (SM). The new physics may become manifest in the form of Higgs boson properties different from those predicted by the SM. One of those properties is expressed by non-standard interactions of the newly discovered $125 \ \text{GeV}$ Higgs-like resonance such as flavor-violating Higgs couplings to leptons and quarks. These interactions could induce non-zero lepton flavor-violating (LFV) Higgs boson decays, such as $h \rightarrow l_i l_j$ with $i \neq j$, indeed the most stringent limits on the branching ratios of LFV  decay of the SM-like Higgs boson Br$(h\rightarrow \mu \tau, e \tau ) < \mathcal{O}(10^{-3})$, from the CMS Collaboration using data collected at a center-of-mass energy of 13 TeV. In contrast, the situation is somewhat more complicated in the quark sector by the process, which is related to the flavor-violating Higgs couplings involving a top quark, which seems to be outside the present reach of LHC. It leads to the experimental upper limits on flavor-changing neutral current (FCNC) decay of top quarks at $95 \%$ CL \cite{FCNC22}. Besides, the strongest indirect bound on FCNC quark--quark--Higgs couplings came from a measurement of meson oscillations. This bound can be translated into the upper bound on the branching fraction of the flavor-violating decay of the Higgs boson to the light quarks \cite{FCNC23}.  

In the physics beyond SM, different mechanisms can yield the non-standard interactions of the SM-like Higgs boson that predict the flavor-violating processes, which could get close to the sensitivities of future accelerators. Among all the possibilities, the models based on the gauge symmetry $SU(3)_C \times SU(3)_L \times U(1)_X \ (3-3-1)$, called the 3-3-1 model \cite{331m, FGN,331r,ecn331,r331},
are rich with FCNC physics, including both quark and lepton sectors \cite{FCNC24, FCNC25}. Besides that, the 3-3-1 model can solve the current issues of physics such as dark matter  \cite{331DM,331DM1, m331}, neutrino mass and mixing \cite{331Neutrino}, the number of fermion generation \cite{FGN}, strong CP conservation \cite{331StrongCP}, electric charge quantization \cite{331charge}. Improved versions of simplifying for solving current experimental results at the larger hadron collider (LHC) have been proposed. Differences in each version make manifest the scalar and fermion contents. The simple 3-3-1 version is an improvement of the minimal and reduced 3-3-1 version \cite{331m,r331}, which contains the smallest fermion and scalar contents \cite{m331}. 
 This improvement allows a simple 3-3-1 model to overcome the disadvantages of the previous models \cite{Simple331m}. The simple 3-3-1 model is realistic on introducing the inner Higgs triplets \cite{m331}. The presence of the inert Higgs triplet not only solves the dark matter problem but also can explain the experimental $\rho$-parameter \cite{m331}. We would like to note that without introducing the inert Higgs triplets, the new physics contribution to the $\rho$-parameter coming from the normal sector is very tiny and can be negligible \cite{DongSi}. However, the inert Higgs triplet gives the contribution to the $\rho$-parameter via a loop effect that is significant, and which is comparable with the global fit, as mentioned in \cite{m331}.
 
 The constraint on the SM-like Higgs boson at the LHC was studied in \cite{m331}. However, one did not consider the implications for collider searches of precision physics bound on the SM-like Higgs bosons with flavor-violating couplings. The simple 3-3-1 model consists of two Higgs triplets in the normal sector and the leptons and quarks couple to both Higgs triplets via general Yukawa matrices (including both normalizable-operators and non-renormalized operators). So, it allows for flavor-changing tree-level couplings of the physical Higgs bosons. It may be able to accommodate large branching ratios for lepton and quark flavor-violating decay of the SM-like Higgs bosons such as $h \rightarrow \mu \tau, h \rightarrow q_i q_j $, with $q_{i,j} $ being a light quark, and the top-quark decays $t \rightarrow q h$. Along with those decays, the decay $\tau \rightarrow \mu \ga$ and the anomalous magnetic moment of the muon $(g-2)_\mu$ also are constrained by the lepton flavor-violating Higgs couplings. The neutral Higgs bosons contribute to $(g-2)_\mu$ at the one-loop level, with both flavors violating vertices, while they contribute to the $\tau \rightarrow \mu \gamma $ at one loop and two loops. 
We hope that these contributions can be fitted to the $(g-2)_\mu$ discrepancy and may reach the current bound on Br$(\tau \rightarrow \mu \gamma)$ of the experiment. So, we are going to focus on studying the contribution of the flavor-violating interactions into some decay channels of the SM-like Higgs boson, heavy quark, and lepton and on $(g-2)_\mu$. 

In Sect.\ref{model}, we briefly review the simple 3-3-1 model. We discuss the constraints from precision flavor observables, such as $h\rightarrow \mu \tau, \tau \rightarrow \mu \gamma$, and $(g-2)_\mu$ in Sect.\ref{Higglepton}. Sect.\ref{Higgquark} investigates  the contributions of flavor violating Higgs couplings to quarks into the meson mixing masses. Based on that research, we show that the branching ratios $h \rightarrow q_i,q_j$, with $i,j \neq 3$ might agree with the upper bound of the experiment. The top-quark decay modes $t \rightarrow q_i h$ also are studied in Sect.\ref{Higgquark}. Finally, we summarize our results and draw conclusions in Sect.\ref{conclusion}.

\section{\label{model}Simple 3-3-1 model}   

The simple model is a combination of the reduced 3-3-1 model~\cite{r331} and the minimal 3-3-1 model \cite{331m} in which the lepton and scalar contents are minimal ~\cite{m331}. The fermion content which is anomaly free is defined as \cite{331m}   
\bea \psi_{aL} &\equiv & \left(\begin{array}{c}
               \nu_{aL}  \\ e_{aL} \\ (e_{aR})^c
\end{array}\right) \sim (1,3,0),\crn 
Q_{\al L}  &\equiv& \left(\begin{array}{c}
  d_{\al L}\\  -u_{\al L}\\  J_{\al L}
\end{array}\right)\sim (3,3^*,-1/3),\hs Q_{3L} \equiv \left(\begin{array}{c} u_{3L}\\  d_{3L}\\ J_{3L} \end{array}\right)\sim
 \left(3,3,2/3\right), \\ u_{a
R}&\sim&\left(3,1,2/3\right),\hs d_{a R} \sim \left(3,1,-1/3\right),\crn
J_{\al R} &\sim&
\left(3,1,-4/3\right),\hs J_{3R} \sim \left(3,1,5/3\right),\nn \eea where $a=1,2,3$
and $\al= 1,2$ are family indices. The quantum numbers in parentheses are given upon assuming 3-3-1
symmetries, respectively. The third generation of quarks  is arranged differently from the two remaining generations to obtain appropriate FCNC contributions when the new energy scale is blocked by the Landau pole. 
Due to the proposed fermion content, the minimal and unique scalars sector is introduced as follows:

  \bea \eta = \left(\begin{array}{c}
	\eta^0_1\\
	\eta^-_2\\
	\eta^{+}_3\end{array}\right)\sim (1,3,0),\hs \chi = \left(\begin{array}{c}
	\chi^-_1\\
	\chi^{--}_2\\
	\chi^0_3\end{array}\right)\sim (1,3,-1),\label{vev2}
\eea with VEVs $ \langle \eta_1^0\rangle = \fr{u}{\sqrt{2}}, \langle \chi_3^0\rangle =\fr{w}{\sqrt{2}}$. In order to reveal candidates for dark matter, an inert scalar multiplet $\phi =\eta^\prime, \chi^\prime$ or $\sigma$, ensured by an extra $Z_2$ symmetry, $\phi \rightarrow - \phi$, is introduced \cite{m331}. Because of $Z_2$ symmetry, the inert and normal scalars do not mix. The physical eigenstates and mass of normal scalars are considered in terms of  $V_{\text{simple}}$ given in \cite{m331}. The Higgs triplets can be decomposed as $\eta^T=(\fr{u}{\sqrt{2}}\ 0\ 0)+(\fr{S_1+i A_1}{\sqrt{2}}\ \eta^-_2\ \eta^+_3)$ and $\chi^T=(0\ 0\ \fr{w}{\sqrt{2}})+(\chi^-_1\ \chi^{--}_2\ \fr{S_3+i A_3}{\sqrt{2}})$. The fields $A_1, A_3$, $\eta_2^\pm, \chi^{\pm \pm}$ and the decomposed state $G_X ^\pm =c_\theta \chi_1^\pm -s_\theta \eta^\pm_3$ are massless Goldstone bosons eaten by $Z$, $Z', W^\pm, Y^{\pm \pm}$, and $X^\pm$  gauge bosons, respectively. The physical scalar fields with respective masses are identified as follows: 
\bea && h \equiv c_\xi S_1-s_\xi S_3,\hs m^2_h=\la_1 u^2+\la_2 w^2-\sqrt{(\la_1 u^2-\la_2 w^2)^2+\la^2_3 u^2 w^2}\simeq \fr{4\la_1\la_2-\la^2_3}{2\la_2}u^2,\crn
&& H \equiv s_\xi S_1 + c_\xi S_3,\hs m^2_{H}=\la_1 u^2+\la_2 w^2+\sqrt{(\la_1 u^2-\la_2 w^2)^2+\la^2_3 u^2 w^2}\simeq 2\la_2 w^2,\\
&& H^{\pm}\equiv c_{\theta}\eta^\pm_3+s_{\theta}\chi^\pm_1,\hs m^2_{H^\pm}=\fr{\la_4}{2}(u^2+w^2)\simeq \fr{\la_4}{2} w^2.\nn \eea
 $\xi$ is the $S_1$--$S_3$ mixing angle, while  $\theta$ is that of $\chi_1$--$\eta_3$ and they are defined via  
$t_\theta=\fr{u}{w},t_{2\xi}=\fr{\la_3 u w}{\la_2 w^2-\la_1 u^2}\simeq \fr{\la_3 u}{\la_2 w}.$ Here, we note that $c_x=\cos(x),\ s_x=\sin(x),\ t_x=\tan(x)$, and so forth, for any $x$ angle. 
The $h$ is identified with the Higgs boson discovered at the LHC and $H$ and $H^\pm$ are new neutral and singly charged Higgs bosons, respectively,

 Because of the conservation of  $Z_2$ symmetry, the inert multiplets do not couple to the fermions. The Yukawa Lagrangian takes the form
  \bea \mathcal{L}_Y=&& h^J_{33}\bar{Q}_{3L}\chi J_{3R}+ h^J_{\al \beta}\bar{Q}_{\al L} \chi^* J_{\beta R}+h^u_{3 a} \bar{Q}_{3L} \eta u_{aR}+ \fr{h^u_{\al a}}{\La} \bar{Q}_{\al L} \eta \chi u_{a R}+ h^d_{\al a} \bar{Q}_{\al L} \eta^* d_{a R} \crn
+ &&\fr{h^d_{3 a}}{\La} \bar{Q}_{3L}\eta^*\chi^* d_{a R}  +h^e_{ab} \bar{\psi}^c_{aL} \psi_{bL}\eta + \fr{h'^e_{ab}}{\La^2}(\bar{\psi}^c_{aL}\eta\chi)(\psi_{bL}\chi^*)+\fr{s^\nu_{ab}}{\La} (\bar{\psi}^c_{aL}\eta^*)(\psi_{bL} \eta^*)+h.c., \label{Yk1}\eea 
where $\La$ is the scale of new physics, which has a mass dimension that defines the effective interactions and needs to yield masses for all the fermions \cite{m331}. Upon the above interactions, the top quark and new quarks obtain masses via renormalization gauge invariant operators while the remaining quarks get masses via non-renormalization gauge invariant operators of dimension $d >4$. After gauge symmetry breaking, a few gauge bosons have mass \cite{m331}. The physical charged gauge bosons with masses are, respectively, given by 
\bea &&W^{\pm}\equiv \fr{A_1\mp i A_2}{\sqrt{2}},\hs m^2_W=\fr{g^2}{4}u^2,\\
&&X^\mp \equiv \fr{A_4 \mp i A_5}{\sqrt{2}},\hs m^2_X=\fr{g^2}{4}(w^2+u^2),\\
&& Y^{\mp\mp}\equiv \fr{A_6 \mp i A_7}{\sqrt{2}},\hs m^2_Y=\fr{g^2}{4}w^2. \eea 
The neutral gauge bosons with corresponding masses are given as follows 
\bea A_\mu && = s_W A_{3\mu}+ c_W\left(-\sqrt{3}t_W A_{8\mu}+\sqrt{1-3t^2_W}B_\mu\right), \hs m_A=0, \\
Z_\mu &&  = c_W A_{3\mu}- s_W\left(-\sqrt{3}t_W A_{8\mu}+\sqrt{1-3t^2_W}B_\mu\right), \hs m_Z^2=\fr{g^2}{4 c_W^2}u^2,\\
Z'_\mu && =\sqrt{1-3t^2_W}A_{8\mu}+\sqrt{3}t_W B_\mu, \hs m^2_{Z^\prime}=\fr{g^2\left[(1-4s_W^2)^2u^2+4c_W^4w^2 \right]}{12c_W^2 (1-4s_W^2)}, \eea
where $\sin \theta_W \equiv s_W=e/g = t/\sqrt{1+4t^2}$, with $t=g_X/g$.
\section{\label{Higglepton}Higgs lepton flavor violating decay}
\subsection{$h\rightarrow$ $\mu \tau$}
Let us consider a non-zero rate for a lepton flavor-violating decay mode of the Higg decay. 
This phenomenology is directly related to the leptonic part of Eq. (\ref{Yk1}). In the physical basis for the scalar, this part can be rewritten as follows:
\bea
\mathcal{L}_Y \supset &&-\bar{e}_{aR}  \left(c_\zeta\fr{1}{u}\left(\mathcal{M}_e \right)_{ab}  -s_\zeta\fr{h^{\prime e}_{ab}}{\sqrt{2}}\fr{uw}{\La^2}  \right)  e_{bL} h-\bar{e}_{aR}  \left(s_\zeta\fr{1}{u}\left(\mathcal{M}_e \right)_{ab}  +c_\zeta\fr{h^{\prime e}_{ab}}{\sqrt{2}}\fr{uw}{\La^2}  \right) e_{bL} H \nonumber  \\ 
&&-\bar{(e_{aL})^c}\left(c_\theta h_{ab}^e  + s_\theta h^{ \prime e}_{ab}\fr{uw}{2\La^2} \right) \nu_{bL}H^+ +\bar{(\nu_{aL})^c}\left(c_\theta h_{ab}^e  \right) e_{bL}H^+ \nonumber \\
&&+\fr{s^{\nu \dag}_{ab}}{\La}\fr{u}{\sqrt{2}}c_\theta\left ( \bar{\nu_{aL}}e_{bR}+\bar{(e_{aR})^c}(\nu_{bL})^c\right)H^++h.c., \label{Yua1}
\eea

where $\left(\mathcal{M}_e \right)_{ab}=\sqrt{2}u\left(h_{ab}^e+\fr{h^{\prime e}_{ab}w^2}{4 \La^2} \right) $ is a mixing mass of charged leptons. We denote 
$e^\prime_{L,R}= \left(e,\mu, \tau \right)_{L,R}=(U^{e}_{L,R})^{-1} \left(e_1,e_2,e_3 \right)_{L,R}$, $\nu_L^\prime=\left(\nu_e, \nu_\mu, \nu_\tau\right )_L=(V^\nu_{L})^{-1}\left (\nu_1, \nu_2,\nu_3 \right)_L$, the Lagrangian given in (\ref{Yua1}) can be rewritten as

\bea
\mathcal{L}_Y \supset && \bar{e}^\prime_Rg_{h}^{ee}e^\prime_L h + \bar{e}^\prime_Rg_H^{ee} e^\prime_LH \nonumber \\ &&+\left\{\bar{(e^\prime_L)^c}g_{L}^{e\nu}\nu^\prime_L+\bar{(\nu^\prime_L)^c}g_{L}^{\nu e}e^\prime_L 
+\bar{\nu^\prime}_L g_R^{\nu e}e^\prime_R +\bar{(e^\prime_R)^c}g_R^{e\nu} (\nu^\prime_L)^c\right\}H^++h.c., \label{HLFV}\eea
where $g_{h}^{ee}=U_{R}^{e\dag}\left(c_\zeta\fr{1}{u}\mathcal{M}_e  -s_\zeta\fr{uw}{\sqrt{2}\La^2} h^{\prime e}\right)U^e_L$,  $g_{H}^{ee}=U_{R}^{e\dag}\left(s_\zeta\fr{1}{u}\mathcal{M}_e +c_\zeta\fr{uw}{\sqrt{2}\La^2} h^{\prime e}\right)U^e_L$, $g_L^{e\nu}=(U_L^e)^T\left(c_\theta  h^e + s_\theta \fr{uw}{2\La^2}h^{e \prime}\right) U_L^\nu$, $g_L^{\nu e}=(U_L^\nu)^T c_\theta h^e U^e_L$, $g_R^{\nu e}=U_L^{\nu \dag}c_\theta \fr{u}{\sqrt{2}\La}s^\nu U^e_R$, $g_R^{e\nu}=U_L^{e T}c_\theta \fr{u}{\sqrt{2}\La}s^\nu U^{\nu T}_R$.

In every parenthesis in the  line of Eq. (\ref{Yua1}), the first term is proportional to the charged lepton masses, whereas the second term in general can contain off-diagonal entries. It is the source of the HLFV processes and leads to the $h \rightarrow e_i e_j$ decays, with $i \neq j$. The branching for this decay process can be written as follows:
 \bea
 \text{Br}(h \rightarrow e_i e_j)= \fr{m_h}{8 \pi \Gamma_h} \left( |g_h^{e_i e_j}|^2+ |g_h^{ e_j e_i}|^2\right),
 \eea
 where $\Gamma_h \simeq 4 \ \text{MeV}$ is the total Higgs boson $h$ decay width, $g_h^{e_ie_j}$ is the Higgs boson $h$ coupling to the charged leptons that we can be obtained from Eq.(\ref{HLFV}). This coupling not only depends on the VEVs and the energy scale $\La$ but also on the Higgs couplings $\la_2, \la_3$. 
 \begin{figure}[h]
 	\resizebox{16cm}{7cm}{\vspace{-2cm}%
 		\includegraphics{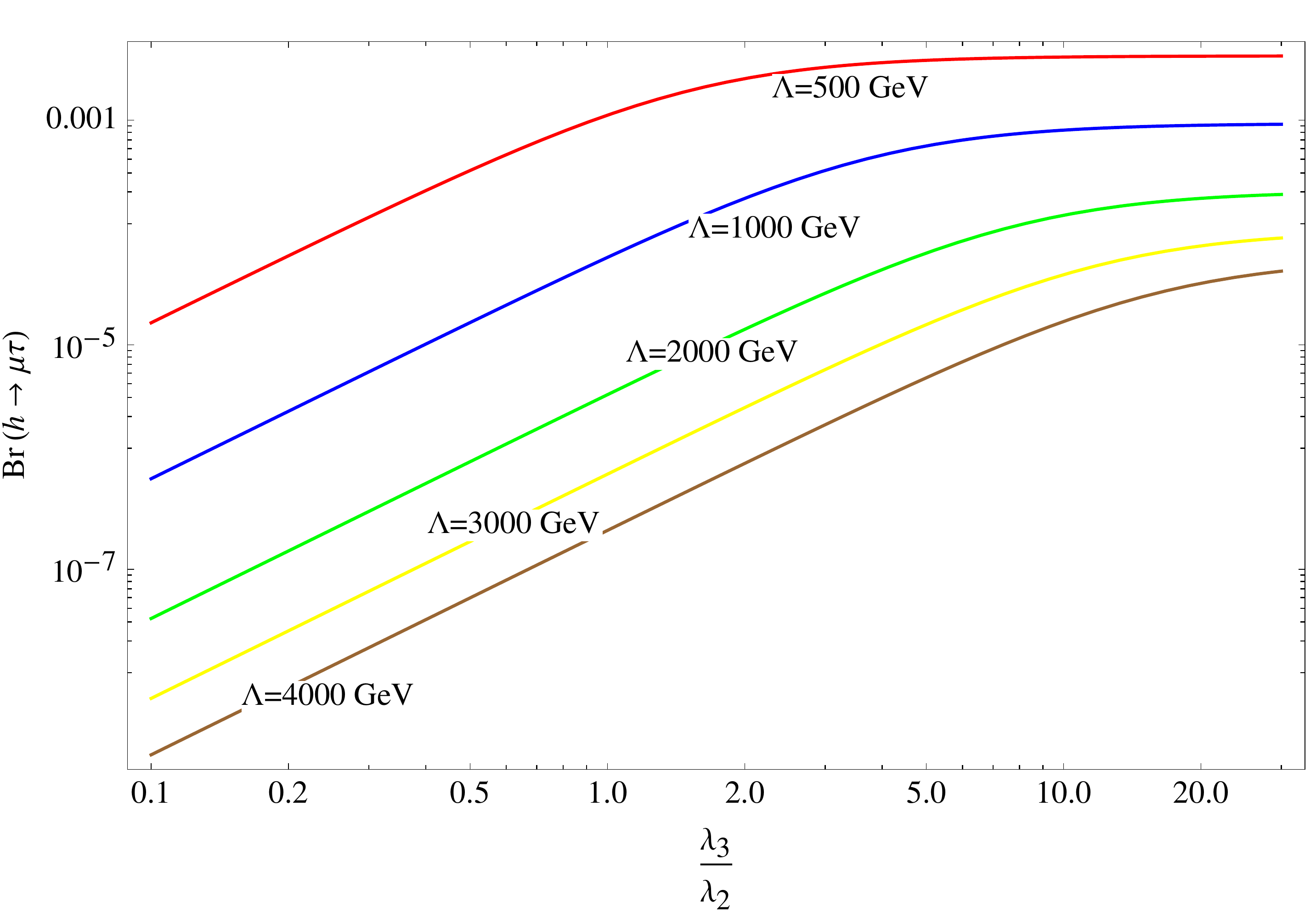}\vspace{0cm}
 		 \includegraphics{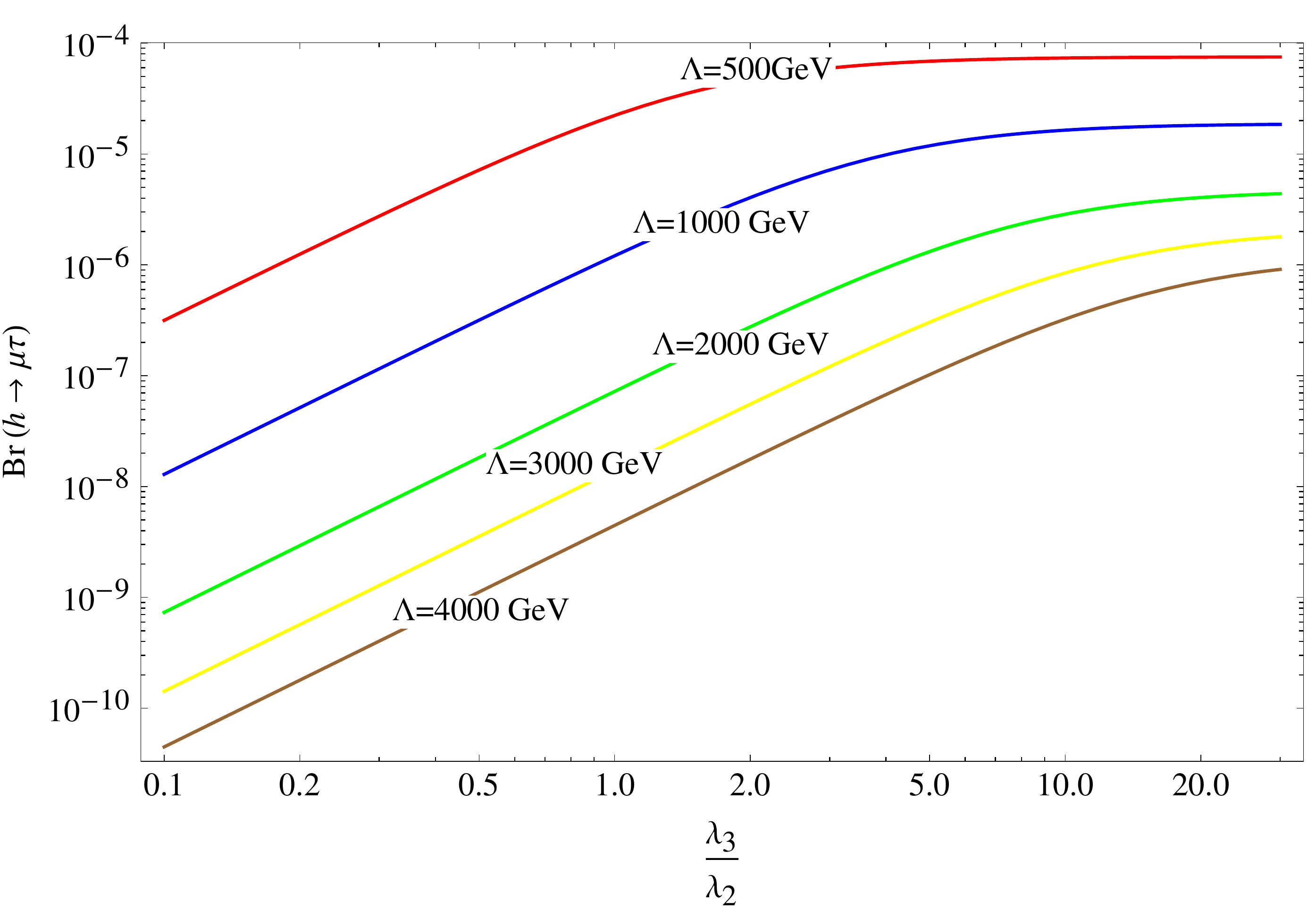}}\vspace{0cm}
 	\caption{The branching ratio Br$(h \rightarrow \mu \tau)$ as a function of factor $\fr{\la_3}{\la_2}$ for different choice of energy scale $\La$. The left and right panels are studied by fixing  $\left[ (U^e_R)^\dag h^{\prime e}U_L^e \right ]_{\mu \tau}=2\fr{\sqrt{m_\mu m_\tau}}{u}$ according to the Cheng--Sher ansatz \cite{Cheng-Sher} and $\left[ (U^e_R)^\dag h^{\prime e}U_L^e \right ]_{\mu \tau}=5 \times 10^{-4}$, respectively  }
 	\label{h-mutau}
 \end{figure}

The numerical result is shown in Fig.\ref{h-mutau} for fixing $u=246 \  \text{GeV}, w=\La$. It is easy to see that the branching ratio of the $h \rightarrow \mu \tau$ can reach the experimental $95 \%$ C.L. upper bounds on the HLFV branching ratios from the CMS Collaborations and also can be as low as $10^{-8}$. It depends quite strongly on the factor $\fr{\la_3}{\la_2}$, $h^{ \prime e}$, and on the energy scale $\La$. In the small $\La$ region and for the factor $\fr{\la_3}{\la_2}>1$, the branching ratio for $h \rightarrow \mu \tau$ can reach $10^{-3}$. However, in this region, the mixing angle of $\xi$ is large. Thus, the simple 3-3-1 model may face stringent constraints such as the Higgs boson couplings to fermions and gauge bosons. If $ \La $ is taken to be a few $\text{TeV}$ but below the Landau pole, $ \la_1, \la_2 $ are of the same order, the mixing angle $\xi$ is small and the branching ratio for $h \rightarrow \mu \tau$ reaches $ 10^{-5} $. 

 \subsection{$\tau \rightarrow \mu \gamma$}
 
 We would like to note that the interaction terms which are given in (\ref{HLFV}) including the lepton flavor-violating and -conserving couplings can affect other LFV processes such as $e_i \rightarrow e_j \gamma$. Besides this contribution, the charged current interactions also induce LFV processes. In the simple 3-3-1 model, the charged current interactions have the following form:
\bea
-\fr{g}{\sqrt{2}}\left( \bar{\nu}_{aL} \ga^\mu e_{aL}W^+_\mu +\bar{\nu}_{aL} \ga^\mu e_{aR}^c X_\mu ^+ +\bar{e}_{aL}\ga^\mu e_{aR}^c Y^{--}_\mu\right)+ h.c..
\eea
Taking all these ingredients into account, the total contribution to the $\tau \rightarrow \mu \gamma$ decay include:
\begin{itemize}
	\item the one-loop diagram with singly charged gauge bosons and neutrinos in the loop	
	\item the one-loop diagram with doubly charged gauge bosons and charged leptons in the loop
	\item the one-loop diagram with charged Higgs bosons and neutrinos in the loop
	\item the one-loop diagram with neutral Higgs bosons and charged leptons in the loop
	\item two-loop Barr-Zee diagrams with an internal photon and a third generation quark
	\item two-loop Barr-Zee diagrams with an internal photon and a gauge boson 
\end{itemize}
The first three types of contributions are the same as those of the 3-3-1 model with a new lepton; see \cite{HueLFV}. The last three types of contributions come from the source of the HLFV processes that are a new contribution and have been not considered in the previous version of the 3-3-1 model \cite{HueLFV}. The total effective Lagrangian describing the $e_i \rightarrow e_j \gamma$ decay process is given as
\bea
e m_\tau \left\{ \bar{e^\prime}_i \left ( D_{R}^\ga \right)_{ij}\sigma^{\al \beta}P_R e_j^\prime F_{\al \beta}
+\bar{e^\prime}_i \left ( D_{L}^\ga \right)_{ij}\sigma^{\al \beta}P_L e_j^\prime F_{\al \beta} \right\}.
\eea
It leads to the branching ratio of the $\tau \rightarrow \mu \gamma$ following processes: 
\bea
\text{Br}(\tau \rightarrow \mu \gamma)=&{}\fr{48 \pi^3 \al}{G_F^2}\left(|D_L^\ga|^2+|D_R^\ga|^2\right)Br(\tau \rightarrow \mu \bar{\nu}_\mu \nu_\tau),
\eea
where $D_{L,R}^\ga$ comes from the one-loop and two-loop diagrams. Firstly, the one-loop diagram contributions with charged Higgs boson $H^{\pm}$, charged gauge boson $W^{\pm}$ and doubly charged gauge boson $Y^{\pm \pm}$ have a form inspired by the general formula in \cite{Lavoura}:
\bea
&& D_{1R}^{\nu W^{\pm}}=-\fr{eg^2m_{\tau}}{32\pi^2 m_W^2}\sum_{j=1}^{3} U^{\nu}_{j 3} U^{\nu *}_{j2}f \left(\fr{m_{\nu_j}^2}{m_W^2}\right), \hs D_{1L}^{\nu W^{\pm}}=-\fr{eg^2m_{\mu}}{32\pi^2 m_W^2}\sum_{j=1}^{3} U^{\nu}_{j 3} U^{\nu *}_{j2}f \left(\fr{m_{\nu_j}^2}{m_W^2}\right), \crn 
&& D_{1R}^{\nu X^{\pm}}=-\fr{eg^2m_{\tau}}{32\pi^2 m_X^2}\sum_{j=1}^{3} U^{\nu}_{j 3} U^{\nu *}_{j2}f \left(\fr{m_{\nu_j}^2}{m_X^2}\right), \hs D_{1L}^{\nu X^{\pm}}=-\fr{eg^2m_{\mu}}{32\pi^2 m_X^2}\sum_{j=1}^{3} U^{\nu}_{j 3} U^{\nu *}_{j2}f \left(\fr{m_{\nu_j}^2}{m_X^2}\right), \crn
&& D_{1R}^{e Y^{\pm \pm}}=-\fr{eg^2m_{\tau}}{32\pi^2 m_{Y^{\pm \pm}}^2}\sum_{j=1}^{3}\left[g\left(\fr{m_{e_j}^2}{m_{Y^{\pm \pm }}^2}\right)-2f \left(\fr{m_{e_j}^2}{m_{Y^{\pm \pm}}^2}\right) \right], \crn 
&&D_{1L}^{e Y^{\pm \pm}}=-\fr{eg^2m_{\mu}}{32\pi^2 m_{Y^{\pm \pm }}^2}\sum_{j=1}^{3}\left[g\left(\fr{m_{e_j}^2}{m_{Y^{\pm \pm }}^2}\right)-2f \left(\fr{m_{e_j}^2}{m_{Y^{\pm \pm }}^2}\right) \right], \crn
&& D_{1R}^{\nu H^{\pm}}=-\fr{eg^2}{32\pi^2 m_{H^{\pm}}^2 m_W^2}\sum_{j=1}^{3} \left\{ g^{\nu \tau *}_L g^{\nu \mu}_L m_{\tau} h \left(\fr{m_{\nu_j}^2}{m_{H^{\pm}}^2}\right) \right. \nonumber \\ && \left. \hs \hs \hs \hs \hs \hs \hs \hs  \hs \hs \hs + g^{\nu \tau *}_R g^{\nu \mu}_R  m_{\mu} k \left(\fr{m_{\nu_j}^2}{m_{H^{\pm}}^2}\right)+g^{\nu \tau *}_L g^{\nu \mu}_R  m_{H} r \left(\fr{m_{\nu_j}^2}{m_{H^{\pm}}^2}\right) \right \}, \crn
&& D_{1L}^{\nu H^{\pm}}=-\fr{eg^2}{32\pi^2 m_{H^{\pm}}^2 m_W^2}\sum_{j=1}^{3}  \left\{ g^{\nu \tau *}_R g^{\nu \mu}_R m_{\tau}h \left(\fr{m_{\nu_j}^2}{m_{H^{\pm}}^2}\right)  \right. \nonumber \\ && \left. \hs \hs \hs \hs \hs \hs \hs \hs  \hs \hs \hs+g^{\nu \tau *}_L g^{\nu \mu}_Lm_{\mu} k \left(\fr{m_{\nu_j}^2}{m_{H^{\pm}}^2}\right)+g^{\nu \tau *}_R g^{\nu \mu}_L  m_{H} r \left(\fr{m_{\nu_j}^2}{m_{H^{\pm}}^2}\right) \right \} . \crn 
\eea
with the functions $f,g,h,k$ and $r$ defined by
\bea
f(x)&=&\fr{10-43x+78x^2-49x^3+4x^4+18x^3\ln{x}}{12(x-1)^4}, \crn
g(x) &=& \fr{8-38x+39x^2-14x^3+5x^4-18x^2\ln{x}}{12(x-1)^4}, \crn
h(x) &=& k(x) = \fr{1-6x+3x^2+2x^3-6x^2\ln{x}}{12(x-1)^4}, \crn
r(x) &=& \fr{-1+x^2-2x\ln{x}}{2(x-1)^3}. 
\eea 
The neutral Higgs contribution to $D_{L,R}^\ga$ via  the one-loop diagram is
\bea
D_{1L}^\ga =D_{1R}^\ga =\sqrt{2} \sum_{\phi}\fr{g_{\phi}^{ \mu \tau}g_{\phi}^{ \tau \tau}}{m_\phi^2} \left( \ln \fr{m_\phi ^2}{m_\tau^2}-\fr{3}{2}\right),
\eea
and the two-loop correction to $D^\ga_{LR}$ being given by \cite{Sacha} 
\bea
&& D_{2L}^{\ga} =  D^{\ga}_{2R} =2 \sum_{\phi ,f} g_{\phi}^{ \mu \tau}g_{\phi}^{ f f} \fr{N_c Q_f^2 \al}{\pi}\fr{1}{m_{\tau}m_f}
f_\phi \left(\fr{m_f^2}{m_\phi^2} \right)\nonumber  \\&&- \sum_{\phi=h, H}g_{\phi}^{ \mu \tau} g_{\phi}^{ GG} \fr{ \al Q_G^2}{2 \pi m_\tau m_G^2}\left\{3f_\phi \left( \fr{m_G^2}{m^2_\phi}\right)+\fr{23}{4}g\left(\fr{m_G^2}{m^2_\phi}\right)+\fr{3}{4}h\left( \fr{m_G^2}{m_\phi^2}\right)+m_\phi^2 \fr{f_\phi \left(\fr{m_G^2}{m_\phi^2}\right)-g\left(\fr{m_G^2}{m_\phi^2} \right)}{2m_G^2} \right\}, \nonumber
\eea
where $\Phi =h, H$, $G= W^\pm, X^\pm, Y^{\pm \pm}$, $f=t,b$, and $Q_G$ is an
electrical charge of the gauge boson $G$. $g_{\phi}^{ \mu \tau}, g_{\phi}^{ f f}, g_{ \phi}^{ G G}$ are the scalar $\phi$ couplings to $\mu$ $\tau$, two fermions, and two gauge bosons $G$, respectively. The expressions for $g_{\phi}^{ f f}, g_{ \phi}^{ G G}$ are given in \cite{m331} and for $g_{\phi}^{ \mu \tau}$ can be obtained from Eq.(\ref{HLFV}).
The loop functions, $f_\phi (z), h(z),g(z)$, are given by \cite{Sacha} 
\bea
f_{h,H}(z)&&=\fr{z}{2}\int_0^1 dx \fr{\left( 1-2x(1-x)\right)}{x(1-x)-z}\ln \fr{x(1-x)}{z}, \nonumber \\
h(z)&&=-\fr{z}{2}\int_0^1 \fr{dx}{x(1-x)-z}\left\{1-\fr{z}{x(1-x)-z}\ln \fr{x(1-x)}{z}\right\} \nonumber \\
g(z)&& =\fr{z}{2}\int^1_0 dx \fr{1}{x(1-x)-z}\ln \fr{x(1-x)}{z}.
\eea 
In the limits $z \gg 1$ and $z \ll 1$, the functions $f(z), g(z)$ and $h(z)$ can approximately be written as follow:
\bea 
z \ll 1, \hs f(z)=\fr{z}{2} (\ln{z})^2, \hs g(z))=\fr{z}{2} (\ln{z})^2, \hs h(z)=z \ln{z}, \crn
z \gg 1 , \hs f(z)=\fr{\ln{z}}{3}+\fr{13}{18}, \hs g(z)=\fr{\ln{z}}{2}+1, \hs h(z)=-\fr{\ln{z}}{2}-\fr{1}{2}.  
\eea 
For $z \sim \mathcal{O}(1)$, the functions $f,g,h \sim z$ can be accurately calculated. Let us estimate each kind of diagrams contributing to $\tau \rightarrow \mu \gamma$ via numerical studies. 
We choose the parameters as follows:
\bea
&& \hs m_W=80.385\  \text{GeV}, \hs m_e=0.000511 \ \text{GeV}, \hs  m_{\mu}=0.1056 \ \text{GeV}, \hs  m_{\tau}=1.176 \ \text{GeV} \crn 
&& \sin^2(\theta_{12})=0.307, \hs \sin^2(\theta_{23})=0.51, \hs \sin^2(\theta_{13})=0.021,\hs \al=\fr{1}{137},  \hs u=246 \ \text{GeV}\crn 
&& \Delta m_{12}^2=m_{\nu_2}^2-m_{\nu_1}^2=7.53 \times 10^{-5} \ \text{eV}^2, \hs \Delta m_{23}^2=m_{\nu_3}^2-m_{\nu_2}^2=2.45  \times 10^{-3} \ \text{eV}^2 \crn &&\la_2=\la_1=0.09, \hs s^{\nu}\sim 10^{-10} . \eea
\begin{figure}[h]
	\begin{tabular}{cc}
		\resizebox{16cm}{7cm}{\vspace{-2cm}%
			\includegraphics{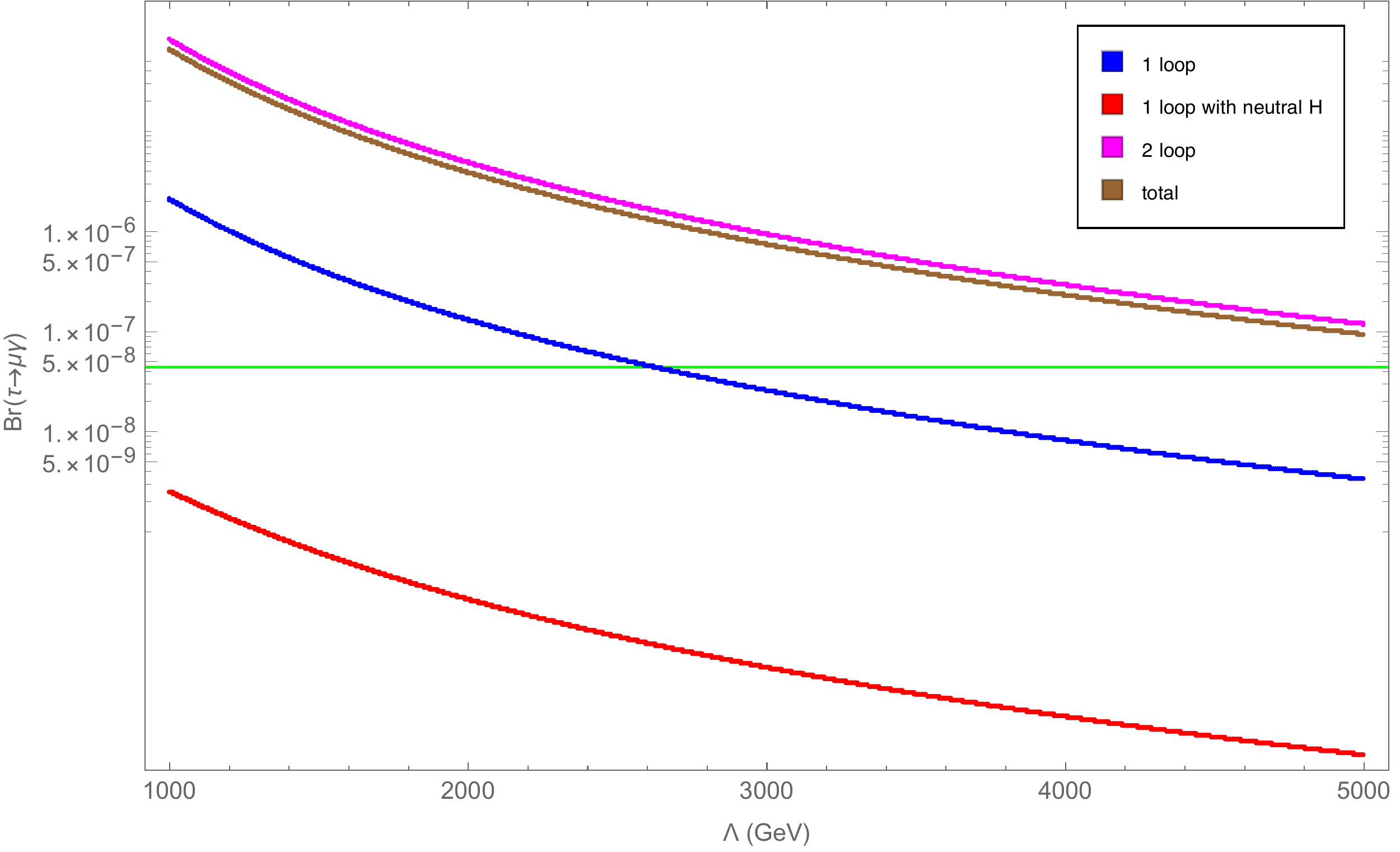}\vspace{0cm}
			\includegraphics{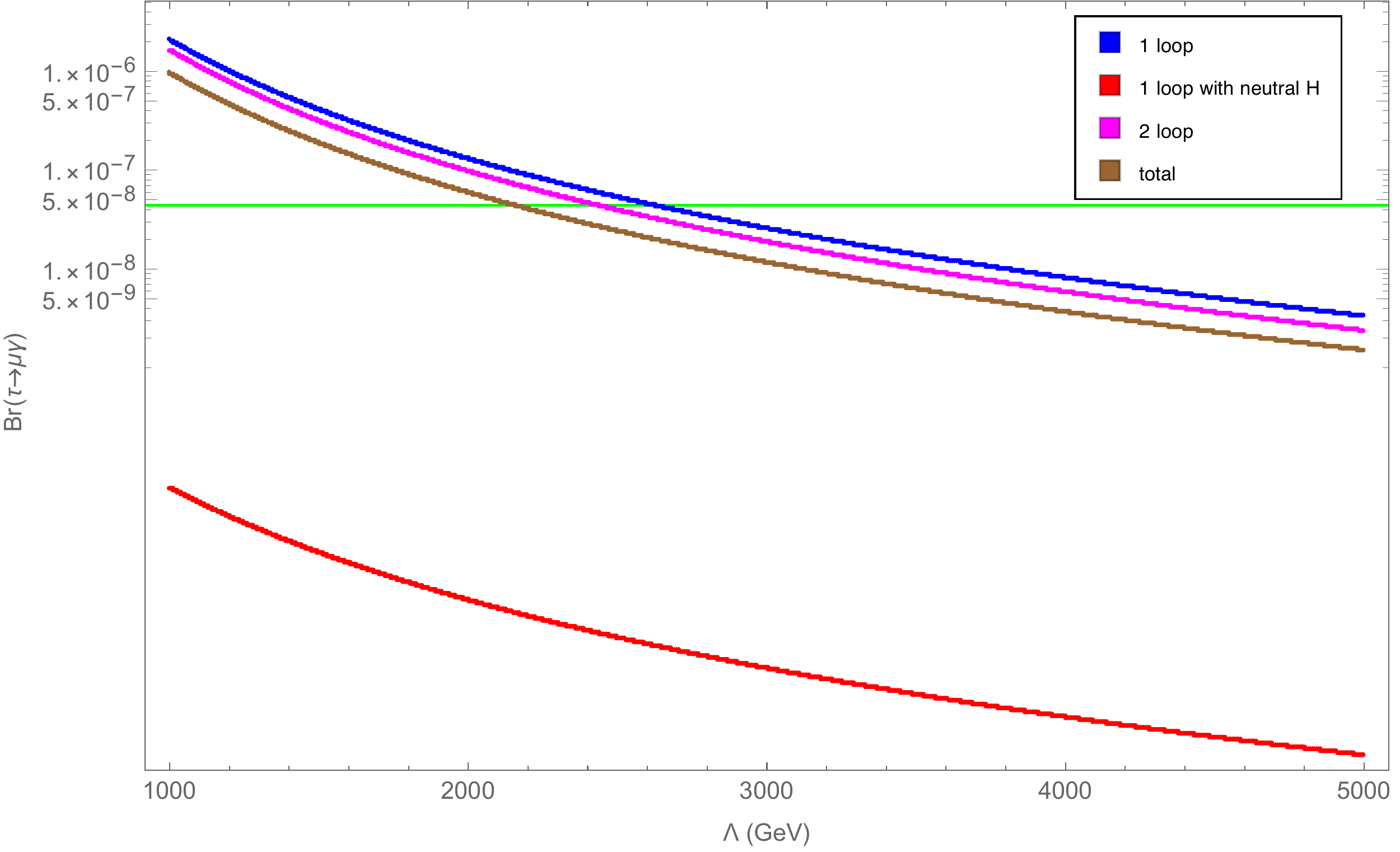}}\vspace{0cm}
	\end{tabular}
	\caption{The dependence of branching ratio Br$(\tau \rightarrow \mu \gamma)$ on the scale of new physics $\La$ in 1-loop, 1-loop with new neutral Higgs boson $H$, 2-loop and total contribution, respectively. The green solid line is the experimental constraint Br$(\tau \rightarrow \mu \gamma)_{\text{Exp}} < 4.4 \times 10^{-8}$. We fix $\left[ (U^e_R)^\dag h^{\prime e}U_L^e \right ]_{\mu \tau}=2\fr{\sqrt{m_\mu m_\tau}}{u}$ and $\left[ (U^e_R)^\dag h^{\prime e}U_L^e \right ]_{\mu \tau}=5 \times 10^{-4}$, for left and right panels, respectively. The factor $\fr{\la_3}{\la_2}=1$ for both panels.    }
	\label{k1}
\end{figure}

\begin{figure}[h]
	\begin{tabular}{cc}
		\resizebox{16cm}{7cm}{\vspace{-2cm}%
			\includegraphics{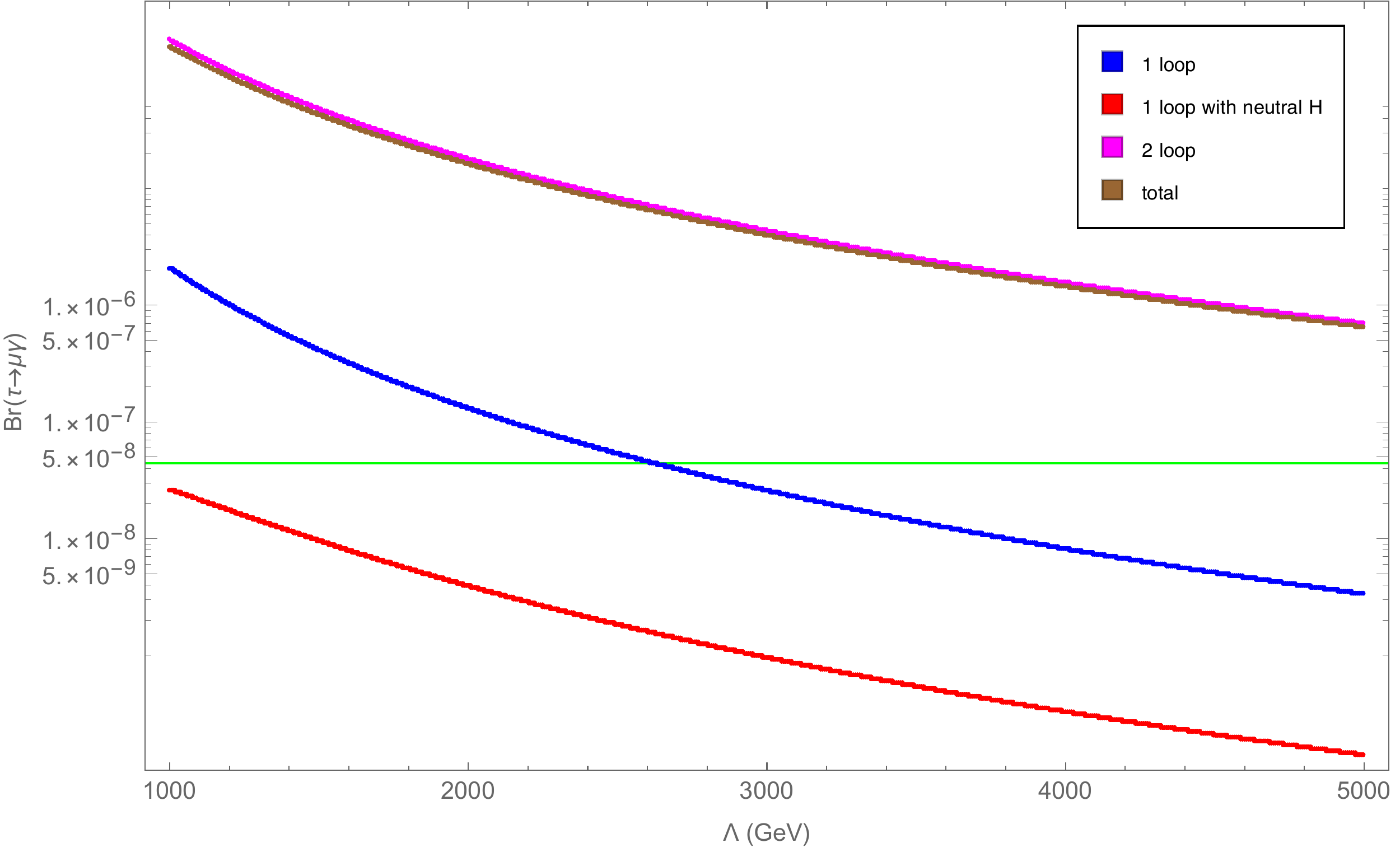}\vspace{0cm}
			\includegraphics{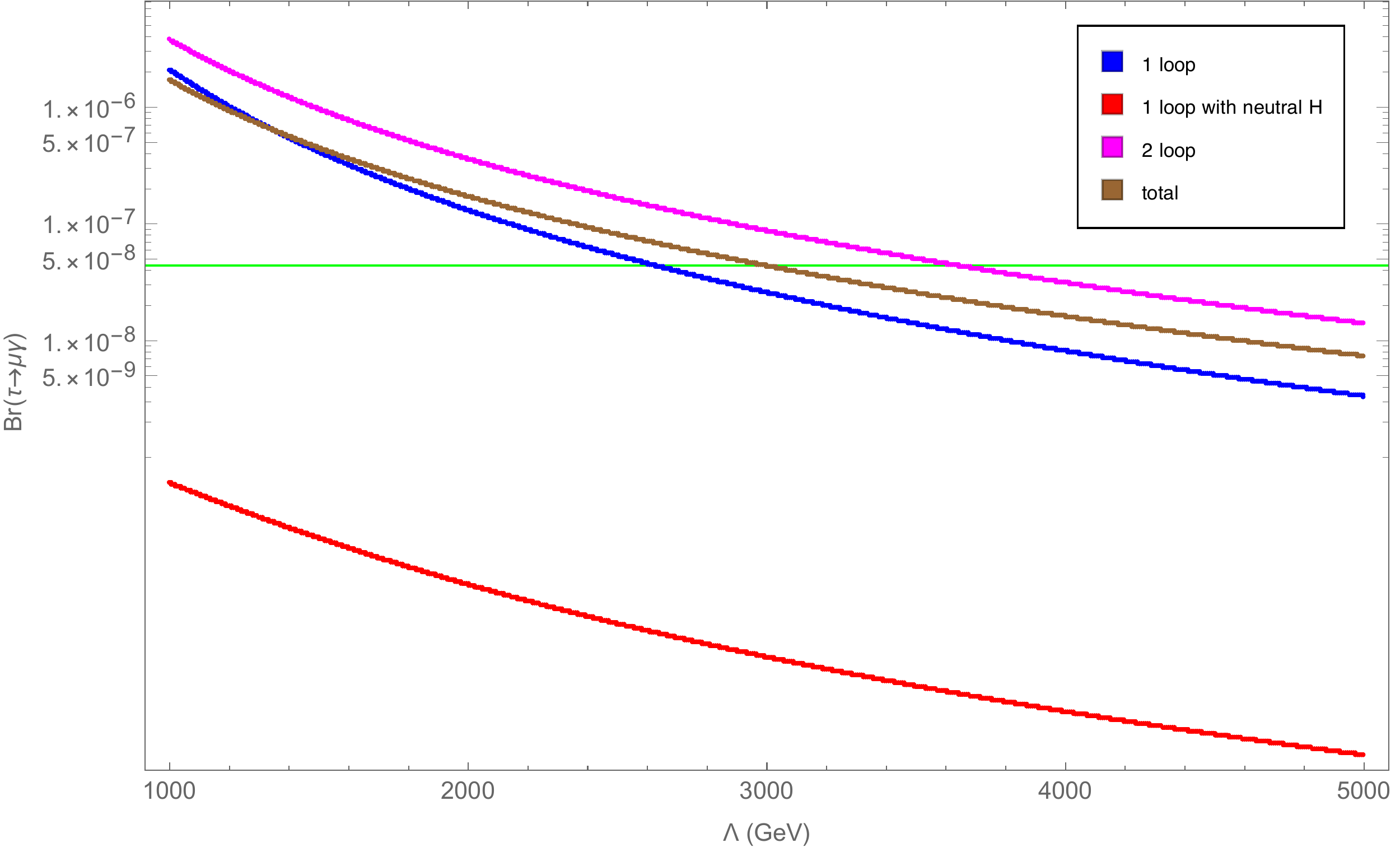}}\vspace{0cm}
	\end{tabular}
	\caption{The dependence of branching ratio Br$(\tau \rightarrow \mu \gamma)$ on the scale of new physics $\La$ in one-loop, one-loop with new neutral Higgs boson $H$, two-loop and total contribution, respectively. The green solid line is the experimental constraint Br$(\tau \rightarrow \mu \gamma)_{\text{Exp}} < 4.4 \times 10^{-8}$. We fix $\left[ (U^e_R)^\dag h^{\prime e}U_L^e \right ]_{\mu \tau}=2\fr{\sqrt{m_\mu m_\tau}}{u}$ by the Cheng--Sher ansatz \cite{Cheng-Sher} and $\left[ (U^e_R)^\dag h^{\prime e}U_L^e \right ]_{\mu \tau}=5 \times 10^{-4}$, for left and right panels, respectively. The factor $\fr{\la_3}{\la_2}=5$ for both panels.   }
	\label{k2}
\end{figure}

The results shown in the  Fig. \ref{k1} suggest that the two-loop diagrams can provide a dominant contribution to $\tau \rightarrow \mu \gamma$. The Br$(\tau \rightarrow \mu \gamma)$ strongly depends on the lepton flavor-violation coupling $\left[ (U^e_R)^\dag h^{\prime e}U_L^e \right ]_{\mu \tau}$. If we choose $\left[ (U^e_R)^\dag h^{\prime e}U_L^e \right ]_{\mu \tau}=2\fr{\sqrt{m_\mu m_\tau}}{u}$, the two-loop contribution to $\tau \rightarrow \mu \ga$ dominates over the one-loop contribution. However, the branch ratio, Br$(\tau \rightarrow \mu \gamma)$, is only consistent with the predictions of the experiment when the new physical scale is above the Landau pole. If $\left[ (U^e_R)^\dag h^{\prime e}U_L^e \right ]_{\mu \tau}=5 \times 10^{-4}$, the two-loop contribution can be less important and the main contribution comes from one-loop diagrams with the conversed lepton couplings. In this case, we obtain the upper bound on the new physics scale: $\La > 2.4 \  \text{TeV}$ from the lower bound on Br$(\tau \rightarrow \mu \ga)$ of the experiment. Comparing the results given in Figs. \ref{k1} and \ref{k2}, we find that the above conclusions change slightly when the factor $\fr{\la_3}{\la_2}$
is changed.

 \subsection{$\left(g-2\right)_\mu$}
The new physics of the 3-3-1 model contributes to the muon's anomalous magnetic moments $a_\mu$ via the flavor conserving
couplings was considered by \cite{BHHL, QRmuon}. The 3-3-1 model also has FCNC, so it can make its own contribution to the anomalous magnetic moment. First, we investigate only the contribution of the FCNC to $(g-2)_\mu$. 
There exists a one-loop contribution to $(g-2)_\mu$ through flavor-violating couplings of the Higgs to $\mu \tau$. According to the work given in \cite{Sacha}, the one-loop contribution mediated by the neutral Higgs contribution to $\left(g-2\right)_\mu$ can be expressed by
\bea
(\Delta a_\mu)^{M331} &&=  \sum_{\phi} \left(g_{\phi}^{ \tau \mu}\right)^2 \fr{m_\mu m_\tau}{8 \pi^2}\int_0^8 dx \fr{x^2}{m_\phi^2-x(m_\phi^2-m_\tau^2)} \nonumber \\ 
&& \simeq \sum_{\phi} \left(g_{\phi}^{ \tau \mu}\right)^2 \fr{m_\mu m_\tau}{8 \pi^2 m_\phi^2}\left( \ln \fr{m_\phi^2}{m_\tau^2} -\fr{3}{2}\right).
\eea
\begin{figure}[h]
	\resizebox{12cm}{8cm}{\vspace{-2cm}%
		\includegraphics{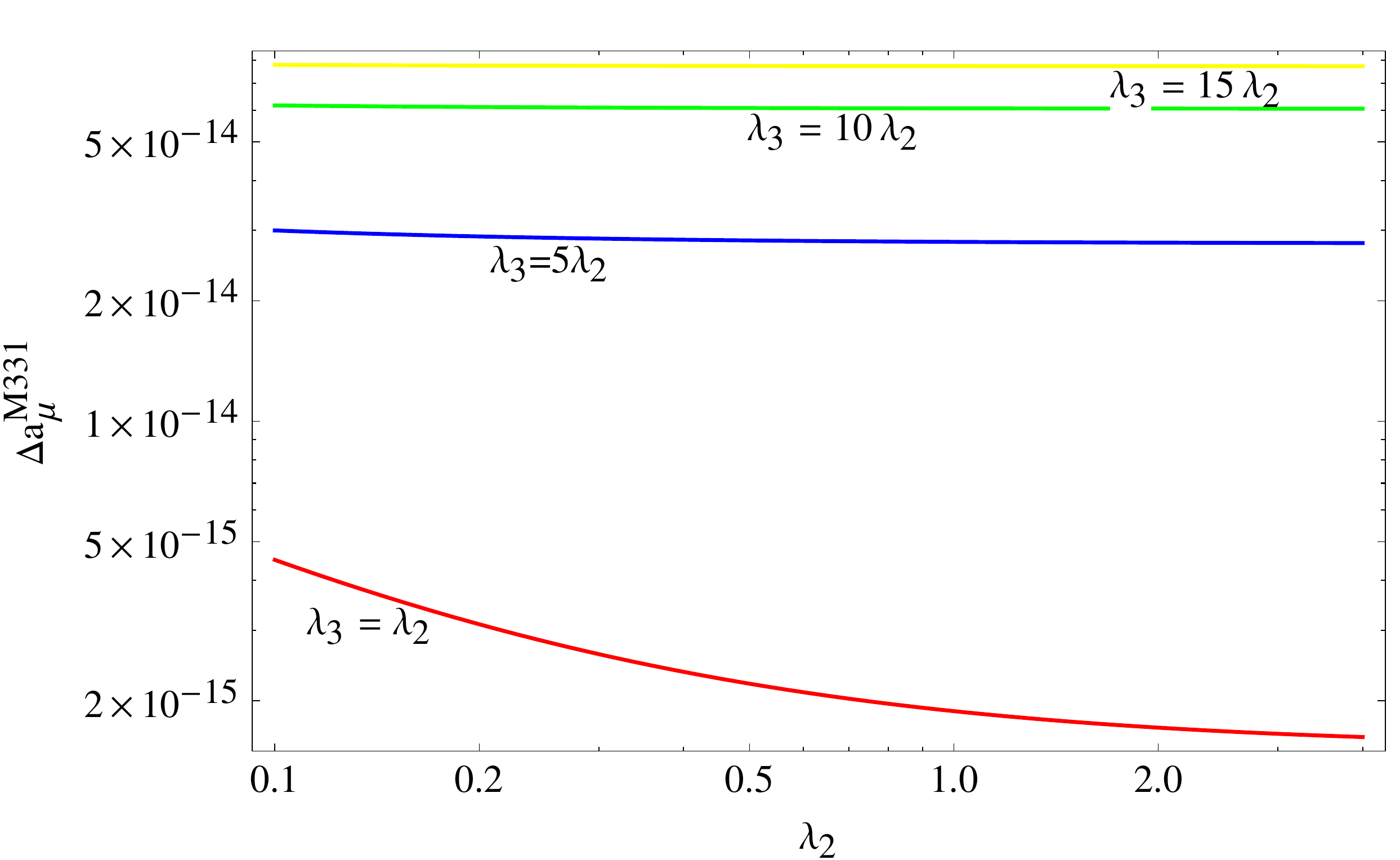}}\vspace{0cm}
	\caption{Contribution of the HLFV interactions to $\Delta a_\mu^{M331}$ as a function of self Higgs coupling $\la_2$ for different factors of $\fr{\la_3}{\la_2}$ and fixing $\La =2000 \ \text{GeV}$.  }
	\label{g-2}
\end{figure}
We plot in Fig. \ref{g-2} the muon's anomalous magnetic moment $\Delta a_\mu^{M331}$ as a function of the self-Higgs coupling $\la_2$, assuming $\La=2000$ GeV, $\left[ (U^e_R)^\dag h^{\prime e}U_L^e \right ]_{\mu \tau}=2\fr{\sqrt{m_\mu m_\tau}}{u}, w=\La, u=246$ GeV. This choice leads to the branching ratio $h\rightarrow \tau \mu$ and can be close to the upper limit value of the experiment or as low as $10^{-5}$ but the flavor-changing interactions of neutral Higgs and two leptons contributed negligibly to $\Delta a_\mu^{M331}$; see Fig. \ref{g-2}. We would like to emphasize that the new contribution to the muon magnetic moment $(g-2)_\mu$ in the context of the simple 3-3-1 model comes from the doubly gauge bosons $Y^{\pm \pm}$, new singly charged vectors $V^\pm$, and new singly charged Higgs $H^\pm$. The dominant contribution is the doubly charged gauge boson running in the loop \cite{ QRmuon}. The total doubly charged boson contribution is given by

\bea
\Delta a_\mu (X^{\pm \pm}) \simeq \fr{28}{3}\fr{m_\mu^2}{u^2+w^2} .
\eea 
It is easy to check that an energy scale of symmetry breaking $SU(3)_L$ around 2 TeV,  1.7 TeV $<w < $2.2 TeV,  is favored as regards explaining the discrepancy of the measured value of muon's anomalous magnetic moment and the one predicted by the standard model \cite{muon},
\bea
(\Delta a_\mu)_{EXP-SM}=(26.1 \pm 8) \times 10^{-10}.
\label{huamuon}\eea
 As mentioned in \cite{m331}, the LHC constraint over the $Z^\prime $ mass in the simple 3-3-1 model is $M_Z^{\prime}> 2.75$ TeV. It can be translated into the lower bound on the VEV, $w$, as follows: $w>2.38 $ TeV. Therefore, the parameter space of w, which is favored for explaining $(\Delta a_\mu)_{EXP-SM}$, is slightly smaller than the lower limit of the LHC (very close to the LHC's allowed space). In other words, in the parameter space that allows an interpretation to be made of LHC's experimental results, the value of the muon's anomalous magnetic moment  is predicted, $(\Delta a_\mu)_{331} < 13.8 \times 10^{-10}$. The upper limit is very close to the constraint given in Eq. (\ref{huamuon}).
\section{\label{Higgquark}Quark flavor-violating Higgs boson decay}

We would like to note that the third family of quarks is transformed differently from the first two families under transformation; it causes the FCNC at the tree level. This works is in \cite{m331}; the authors studied the tree-level FCNCs due to the new neutral gauge boson exchange. However, the FCNC is not only caused by the new neutral gauge boson ($Z'$) exchange but also caused by the SM Higgs boson and a new Higgs boson.
After electroweak symmetry breaking, the operators of Eq.(\ref{Yk1}) give rise to the interaction of neutral Higgs bosons with a pair of SM quark of the form
\bea
\mathcal{L}_Y && \supset -\bar{u}_{aR}\left\{ c_\xi \fr{1}{u} \left(M^u \right)_{ab}+s_\xi \fr{h_{ab}^u}{\La}\fr{u}{2}\right\} u_{bL}h -\bar{u}_{aR} \left\{s_\xi \fr{1}{u} \left(M^u \right)_{ab}-c_\xi \fr{h_{ab}^u}{\La}\fr{u}{2} \right\} u_{bL}H \nonumber \\
&&-\bar{d}_{aR} \left\{c_\xi \fr{1}{u} \left(M^d \right)_{ab}-s_\xi \fr{h_{ab}^d}{\La}\fr{u}{2} \right\} d_{bL}
-\bar{d}_{aR} \left\{s_\xi \fr{1}{u} \left(M^d \right)_{ab}+c_\xi \fr{h_{ab}^d}{\La}\fr{u}{2} \right\} d_{bL}H+h.c.,
\label{LYq}\eea 
where $h^u_{ab}=0$ if $a = 3$, $h^d_{ab}=0$ if $a = 1,2$ and the remaining values of $h_{ab}^u, h_{ab}^d$ are nonzero. We define the physical sates $u_{L,R}^\prime = (u_{1L,R}^\prime, u_{2L,R}^\prime, u_{3L,R}^\prime)^T$, 
$d_{L,R}^\prime = (d_{1L,R}^\prime, d_{2L,R}^\prime, d_{3L,R}^\prime)^T$. They are related to the flavor states $u=(u_{1L,R},u_{2L,R},u_{3L,R})^T, d=(d_{1LR},d_{2LR},d_{3L,R})^T$ by $V_{L,R}^{u,d}$ matrices by
$u_{L,R}=V_{L,R}^u u_{L,R}^\prime, d_{L,R} =V_{L,R}^d d_{L,R}^\prime$.  In the physical states, the Lagrangian given in Eq.(\ref{LYq}) can be rewritten as follows:
\bea
\mathcal{L}_Y && \supset \bar{u^\prime}_{R}\mathcal{G}_h^u u^\prime_{L}h+\bar{d^\prime}_{R}\mathcal{G}_h^d 
 d^\prime_{L} h+\bar{u^\prime}_{R}\mathcal{G}_H^u u^\prime_{L}H+\bar{d^\prime}_{L}\mathcal{G}^d_H 
 d^\prime_{R}H+h.c., \label{FLCN1}
\eea 
where $\mathcal{G}^u_h= -\left(V^u_{R}\right)^\dag \left\{ c_\xi \fr{1}{u} M^u +s_\xi \fr{h^u}{\La}\fr{u}{2}\right\} V^u_L $,  $\mathcal{G}^d_h= -\left(V^d_{R}\right)^\dag \left\{ c_\xi \fr{1}{u} M^d -s_\xi \fr{h^d}{\La}\fr{u}{2}\right\} V^d_L$, $\mathcal{G}^u_H= -\left(V^u_{R}\right)^\dag \left\{ s_\xi \fr{1}{u} M^u-c_\xi \fr{h^u}{\La}\fr{u}{2}\right\} V^u_L $, and $\mathcal{G}^d_H=-\left(V^d_{R}\right)^\dag \left\{ s_\xi \fr{1}{u} M^d +c_\xi \fr{h^d}{\La}\fr{u}{2}\right\} V^d_L.$

Besides, the tree-level FCNC associated with the field $Z^\prime_\mu$ is given in \cite{m331} as
\bea
\mathcal{L}_{FCNC}=-\fr{g}{\sqrt{3}\sqrt{1-3t_W^2}}\left\{\left( V^*_{qL}\right)_{3i}\left( V_{qL}\right)_{3j}\bar{q}^\prime_{iL} \ga^\mu q^\prime_{jL} Z^\prime_\mu\right\}. \label{FNCNZ}
\eea
We would like to recall that the tree-level FCNC associated with the neutral gauge boson $Z^\prime$ was considered in \cite{m331}. The strongest bound on the $Z^\prime$ mass, $m_{Z^\prime}> 4.67$ TeV, came from a measurement of $B_s$--$\bar{B}_s$
oscillations. This value is close to a Landau pole. Around this point, the gauge coupling of the $U(X)$ becomes very large and thus the theory loses the perturbative character. To avoid this difficulty, we extinguish the tree-level FCNC source caused by the new gauge boson $Z^\prime$ in the d-quark sector by setting $\left( V_{dL}\right)_{3a}=0$. Therefore, only the flavor-violating Higgs couplings to quarks can generate the FCNC at tree level, and these couplings can be constrained by $K^0$ and $B^0_{s,d}$ meson oscillation experiments. After integrating out the Higgs fields, the effective Lagrangian for meson mixing can be written as follows:

\bea
\mathcal{L}_{FCNC}^{eff}&&=  \left\{\frac{\left[\left(\mathcal{G}_h^q\right)_{ij} \right]^2}{m_h^2}+\frac{\left[\left(\mathcal{G}_H^q\right)_{ij} \right]^2}{m_H^2}\right\}\left(\bar{q}_{iR}q_{jL}\right)^2  + \left\{\frac{\left[\left(\mathcal{G}_h^q\right)_{ji}^* \right]^2}{m_h^2}+\frac{\left[\left(\mathcal{G}_H^q\right)_{ji}^* \right]^2}{m_H^2}\right\}\left(\bar{q}_{iL}q_{jR}\right)^2 \nonumber \\ && +2\left\{\frac{\left[\left(\mathcal{G}_h^q\right)_{ij} \right]}{m_h}+\frac{\left[\left(\mathcal{G}_H^q\right)_{ij} \right]}{m_H}\right\}\left\{\frac{\left[\left(\mathcal{G}_h^q\right)_{ji}^* \right]}{m_h}+\frac{\left[\left(\mathcal{G}_H^q\right)_{ji}^* \right]}{m_H}\right\}\left(\bar{q}_{iR}q_{jL}\right)\left(\bar{q}_{iL}q_{jR}\right).
\label{FNCN1}\eea
The predicted results for $B_{d,s}$--$\bar{B}_{d,s}$, $K^0$--$\bar{K^0}$, and $D^0$--$\bar{D}^0$ mixing are obtained as in \cite{DHLN}.  Note that there are two scalar fields that have flavor-violating couplings to quarks. Both of them yield the FCNC at tree level.
To compare the contribution of each type, let us estimate the ratio $\kappa\equiv\frac{\left [\left( \mathcal{G}_h^q\right)_{ij}\right]^2 m_H^2}{\left( \left[\mathcal{G}_H^q \right)_{ij}\right]^2 m_h^2} \simeq  \fr{m_H^2}{m_h^2} \tan^2 \xi$. In the limit, $w>>u$, we find the value of $\kappa $ to be always smaller than one unit.  This means that the new scalar Higgs gives more contributions to the FCNC than the SM like Higgs boson. Fitting these results with the experimental measurements of $\Delta m_{B_{s,d}}, \Delta m_{D}, \Delta m_{K^0}$, we get the bound on the flavor-violating Higgs couplings. The strongest bound for new physics comes from the $B_s$--$\bar{B}_s$ mixing. The experimental values of $\Delta m_{B_{s}}$ lead to the bound on $(\mathcal{G}_h^q)_{32}$ as follows:
\bea
2\left(1+\frac{1}{\ka}\right) |\left(\mathcal{G}_h^q\right)_{32}|^2=2\left(1+\frac{1}{\ka}\right)\frac{\la_3^2 u^4}{\la_2^2 w^4}|\left[ (V_R^d)^\dag h^d V_{L}^d\right]_{23}|^2 <1.8 \times 10^{-6}. \label{newphysic}
\eea
The lower bound on the new physics scale $w$ depends on the choice of other parameters. Due to $\frac{\la_3}{\la_2}>1$ and $V_R^d, h^d$ not being fixed,  the constraints from the mixing mass matrix of the mesons not only translate to the new physics scale, $w$, but also translate to other parameters. Therefore, the new physics scale can be chosen far from the Landau pole. The perturbative character of the theory is ensured. 

The constraints on the flavor-violating SM-like Higgs boson couplings to the quarks can be translated into 
upper limits on the branching fraction of the flavor-violating decays of the SM like Higgs boson to light quarks.
In our model, the upper limits for the branching ratios of $h\rightarrow q_iq_j$ are predicted to decrease by $\frac{1}{1+\frac{1}{k}}$ times that of the predictions in \cite{DeltaB}; for details see Table \ref{constrainthdecy}. The weakest constraints are in the $b$--$s$ sector, Br$(h- b \bar{s}) < 3.5\times 10^{-3}$, which is too small to be observed at the LHC because of the large QCD backgrounds, but these signals are expected to be observed at the ILC \cite{ILC} in the future.\\
\begin{table}
\begin{tabular}{|m{10em}|m{10em}|}
	\hline 
	Observable  & Constraint\\ 
	\hline 
	$D^0$ oscillations & Br$(h \rightarrow u \bar{c}) \leq \frac{2 \times 10^{-4}}{1+\frac{1}{\ka}}$\\ 
	\hline 
		$B_d^0$ oscillations & Br$(h \rightarrow d \bar{b}) \leq \frac{8 \times 10^{-5}}{1+\frac{1}{\ka}}$\\
		\hline
			$K^0$ oscillations & Br$(h \rightarrow d \bar{s}) \leq \frac{2 \times 10^{-6}}{1+\frac{1}{\ka}}$\\
		\hline	$B_s^0$ oscillations & $Br(h \rightarrow s \bar{b}) \leq \frac{7 \times 10^{-3}}{1+\frac{1}{\ka}}$\\
		\hline
\end{tabular}
{\caption{The upper limit on  flavor-violating decays of the SM-like Higgs boson to the light quarks at $95 \%$ CL from experiments with mesons }.\label{constrainthdecy}}
\end{table}\\

\begin{figure}[h]
	\resizebox{12cm}{12cm}{\vspace{-2cm}%
		\includegraphics{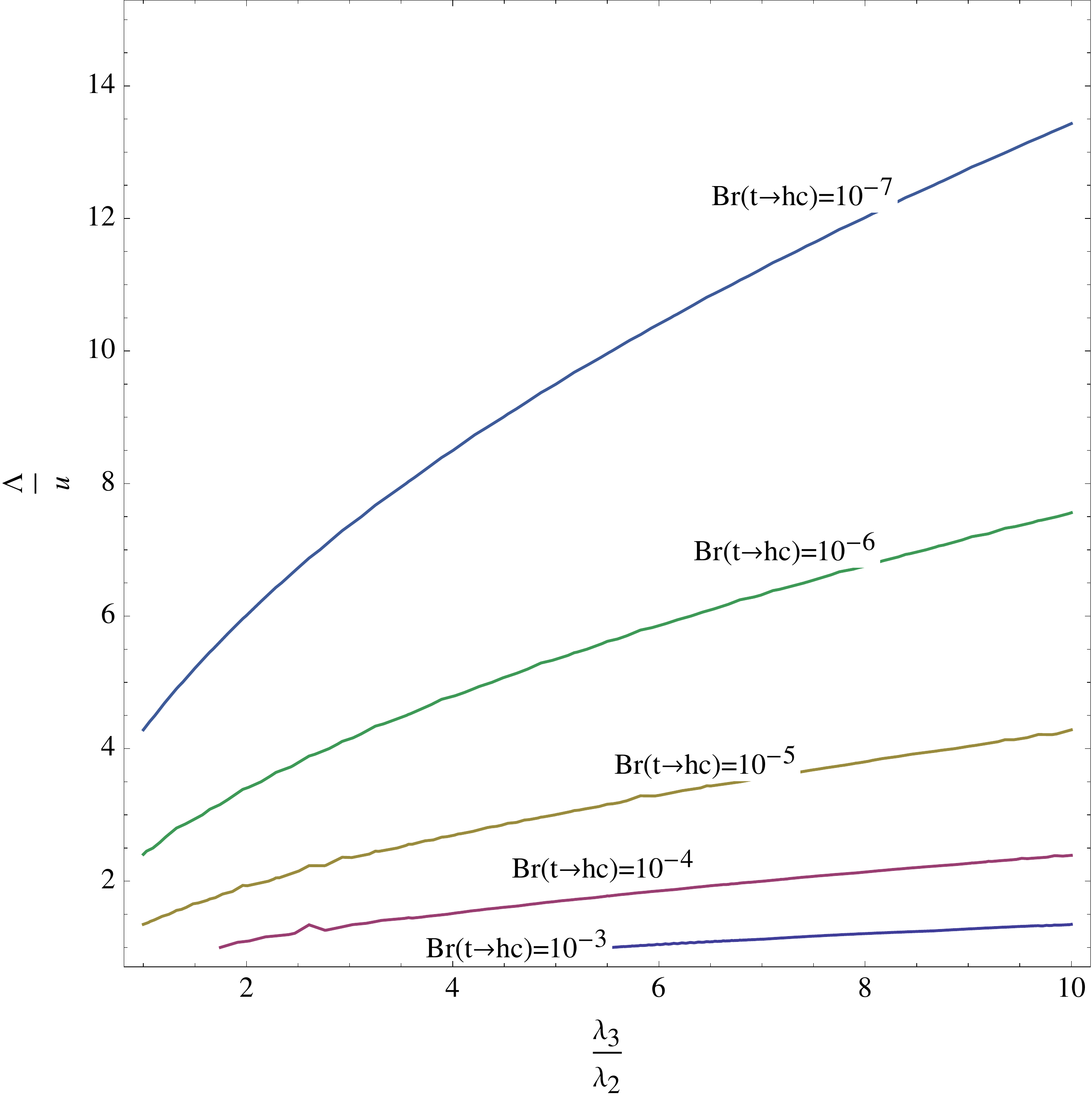}}\vspace{0cm}
	\caption{Top-quark rare decays into $hc$. }
	\label{t-hc}
\end{figure}

The flavor-violating Higgs couplings to the quarks given in Eq.(\ref{FLCN1}) leads to the non-standard top-quark decay mode $t \rightarrow hu_i$, $u_i= u,c$, the rate for which is given by (here we have neglected terms of  $\mathcal{O}(\fr{m_c^2}{m_t^2})$)
\bea
\Gamma(t \rightarrow h u_i)= \fr{|\mathcal{G}^u_{i3}|^2+|\mathcal{G}^u_{3i}|^2}{16\pi}\fr{\left(m_t^2-h_h^2\right)^2}{m_t^3}.
\eea
The branching ratio for the decay $t \rightarrow u_i h$ is defined as follows
\bea
\text{Br}(t \rightarrow u_i h) \simeq \fr{\Gamma(t \rightarrow u_i h)}{\Gamma(t \rightarrow b W^+)}, 
\eea
where $\Gamma(t \rightarrow b W^+) \simeq \fr{g^2 m_t}{64 \pi} \left(1-\fr{m_W^2}{m_t^2}\right)\left(1-2\fr{m_W^2}{m_t^2}+\fr{m_t^2}{m_W^2}\right)$.
The Higgs mediated FCNC in top-quark sector is actively investigated at the LHC by \cite{t-FCNC}. No signal is observed and the upper limit on the branching fractions Br$(t \rightarrow h c) < 0.16 \%$ and Br$(t \rightarrow h u) < 0.19 \%$ at $95 \%$ confidence level are obtained. In Fig. \ref{t-hc}, we plot the Br$(t\rightarrow hc)$ in the  $\fr{\la_2}{\la_3}$--$\fr{w}{u}$ plane for fixing $\left[\left(V_{R}^u\right)^\dag h^u V_{L}^u \right]_{32}=\left[\left(V_{R}^u\right)^\dag h^u V_{L}^u \right]_{23}=2\fr{\sqrt{m_c m_t}}{u}$. The top-quark rare decays into $hc$ could reach up to $10^{-3}$ if the new physics scale is several hundred GeVs, and the factor $\frac{\la_3}{\la_2} > 5$. In this region of parameter space, the mixing angle $\xi$ is large. The model encounters the difficulties as mentioned in the Sect. \ref{Higgquark}. Br$(t \rightarrow ch)$ drops rapidly as the factor $\fr{w}{u}$ is increased. For small mixing angle $\xi$, Br$(t \rightarrow hc )$ varies from $10^{-5}$ to $10^{-8}$. Our results are consistent with \cite{t-FCNC1}. 

\section{\label{conclusion} Conclusion}
We study the non-standard interactions of the SM-like Higgs boson that allows for sizable effects in FCNC processes in the simple 3-3-1 model. We examine some effects in flavor physics and constraints on the model both from the quark and lepton sectors via renormalizable and non-renormalizable Yukawa interactions. Specifically, due to the couplings of the leptons to both Higgs triplets, it creates the lepton flavor-violating couplings at tree level. The existence of these interactions is completely independent of the source of non-zero neutrino masses and mixing. This means that, if the source generating mass for neutrinos is turned off, the lepton flavor-violation processes such as $h \rightarrow l_i l_j$ or $l_i \rightarrow l_j \gamma$... are perfectly possible. The branching for $h \rightarrow \mu \tau$ decay depends on the non-renormalizable Yukawa coupling $h^{\prime e}$,  the mixing angle $\xi$, and the new physical scale. For large mixing angle $\xi$, Br$(h \rightarrow \mu \tau)$ can reach the experimental upper bound of the ATLAS and CMS, while for small mixing angle, Br$(h \rightarrow \mu \tau)$ can be $10^{-5}$. The $\tau \rightarrow \mu \gamma$ radiative decay is considered by both lepton flavor-conserving and -violating couplings. The contributions coming from two-loop diagrams with lepton flavor-violating vertex and all one-loop diagrams (including lepton flavor-violating/conserving vertexes) are comparable.  The lepton flavor-violating contribution to $(g-2)_\mu$ is suppressed if the  parameters are selected to satisfy the limits from $h\rightarrow \mu \tau$ and $\tau \rightarrow \mu \gamma$, while the flavor-conserving coupling of the muon to the new gauge boson $Y_\mu ^{\pm \pm}$ almost allows one to explain the muon's anomalous magnetic moment, $(\Delta a_\mu)_{331} < 13.8 \times 10^{-10}$, due to the constraint of LHC over the $Z^\prime$ mass.

We investigate the flavor-violating interactions of the Higgs boson with a pair of quarks. These interactions not only generate the FCNC, which are testable in $B_{s,d}, K^{0}$ meson oscillation experimental but also introduce the additional decay modes for the Higgs boson. The binding conditions from the meson oscillation experiment were transferred to the upper limit on the branching ratio of these decay. They are $\fr{1}{1+\frac{1}{\kappa}}$ smaller than the predictions given in \cite{DeltaB}.Directly testing these Higgs decays seems to be outside the LHC reach but
they are promising as regards searching in the future ILC \cite{ILC}. A search for FCNC in events with the top quark is presented. The upper bound on the branching fraction of top-quark decay, $t \rightarrow h c$ strongly depends on the new physics scale. It can reach $10^{-5}$ or be as low as $10^{-8}$.  
\section*{Acknowledgments}\vspace{-0.4cm}
This research is funded by the Vietnam National Foundation for Science and Technology Development (NAFOSTED) 
under Grant number 103.01-2019.312.

\appendix

\end{document}